# No supernovae associated with two long-duration γ-ray bursts

Johan P. U. Fynbo[1], Darach Watson[1], Christina C. Thöne[1], Jesper Sollerman[1,2], Joshua S. Bloom[3], Tamara M. Davis[1], Jens Hjorth[1], Páll Jakobsson[4], Uffe G. Jørgensen[5], John F. Graham[6], Andrew S. Fruchter[6], David Bersier[7], Lisa Kewley[8], Arnaud Cassan[9], José María Castro Cerón[1], Suzanne Foley[10], Javier Gorosabel[11], Tobias C. Hinse[5], Keith D. Horne[12], Brian L. Jensen[1], Sylvio Klose[13], Daniel Kocevski[3], Jean-Baptiste Marquette[14], Daniel Perley[3], Enrico Ramirez-Ruiz[15,16], Maximilian D. Stritzinger[1], Paul M. Vreeswijk[17,18], Ralph A. M. Wijers[19], Kristian G. Woller[5], Dong Xu[1], Marta Zub[6]

[1]*Dark Cosmology Centre, Niels Bohr Institute, University of Copenhagen, Juliane Maries Vej 30, DK-2100 Copenhagen, Denmark*

[2]*Department of Astronomy, Stockholm University, Sweden*

[3]*Department of Astronomy, University of California at Berkeley, 601 Campbell Hall, Berkeley, CA 94720, USA*

[4]*Centre for Astrophysics Research, University of Hertfordshire, College Lane, Hatfield, Herts, AL10 9AB, UK*

[5]*Niels Bohr Institute, University of Copenhagen, Juliane Maries Vej 30, DK-2100 Copenhagen, Denmark*

[6]*Space Telescope Science Institute, 3700 San Martin Drive, Baltimore, Maryland 21218, USA*

[7]*Astrophysics Research Institute, Liverpool John Moores University, Twelve Quays House, Egerton Wharf, Birkenhead CH41 1LD, UK*

[8]University of Hawaii, Institute of Astronomy, 2680 Woodlawn Drive, Honolulu, HI 96822, US

[9]Astronomisches Rechen-Institut (ARI), Zentrum für Astronomie der Universität Heidelberg (ZAH), Mönchof Str. 12-14, 69120 Heidelberg, Germany

[10]School of Physics, University College Dublin, Dublin 4, Ireland

[11]Instituto de Astrofísica de Andalucía (CSIC), Apartado de Correos, 3004, E-18080 Granada, Spain

[12]SUPA Physics/Astronomy, University of St. Andrews, St. Andrews KY16 9SS, Scotland, UK

[13]Thüringer Landessternwarte Tautenburg, Sternwarte 5, D-07778 Tautenburg, Germany

[14]Institut d'Astrophysique de Paris, UMR70951 CNRS, Université Pierre & Marie Curie, 98 bis boulevard Arago, 75014 Paris, France

[15]Institute for Advanced Study, Einstein Drive, Princeton, NJ 08540, USA

[16]Department of Astronomy and Astrophysics, University of California, Santa Cruz, CA 95064, USA

[17]European Southern Observatory, Alonso de Córdova 3107, Casilla 19001, Vitacura, Santiago, Chile

[18]Departamento de Astronomía, Universidad de Chile, Casilla 36-D, Santiago, Chile

[19]Astronomical Institute 'Anton Pannekoek', Faculty of Science, University of Amsterdam, Kruislaan 403, 1098 SJ Amsterdam, the Netherlands

**It is now accepted that long duration γ-ray bursts (GRBs) are produced during the collapse of a massive star[1,2]. The standard "collapsar" model[3] predicts that a broad-lined and luminous Type Ic core-collapse supernova (SN) accompanies every long-duration GRB[4]. This association has been confirmed in observations of several nearby GRBs [5–9]. Here we present observations of two nearby long-duration GRBs that challenge this simple view. In the GRBs 060505 and 060614 we demonstrate that no SN emission accompanied these long-duration bursts[10,11], down to limits hundreds of times fainter than the archetypal SN 1998bw that accompanied GRB 980425, and fainter than any Type Ic SN ever observed[12]. Multi-band observations of the early afterglows, as well as spectroscopy of the host galaxies, exclude the possibility of significant dust obscuration and show that the bursts originated in actively star-forming regions. The absence of a SN to such deep limits is qualitatively different from all previous nearby long GRBs and suggests a new phenomenological type of massive stellar death.**

The GRBs 060505 and 060614 were detected by the γ-ray Burst Alert Telescope (BAT) onboard the dedicated GRB satellite *Swift* on 2006 May 5.275 and 2006 June 14.530 respectively[11,12]. GRB 060505 was a faint burst with a duration of 4 s. GRB 060614 had a duration of 102 s and a pronounced hard to soft evolution. Both were rapidly localised by *Swift*'s X-ray telescope (XRT). Subsequent follow-up of these bursts led to the discovery of their optical afterglows, locating them in galaxies at low redshift: GRB 060505 at $z = 0.089$[13] and GRB 060614 at $z = 0.125$[14,15]. The relative proximity of these bursts engendered an expectation that a bright SN would be discovered a few days after the bursts, as had been found just a few months before in another low-redshift *Swift* burst, GRB 060218 ($z = 0.033$)[9], and in all previous well-observed nearby bursts[1,5-8].

We monitored the afterglows of GRB 060505 and 060614 using a range of telescopes (see supplementary material for details). These led to early detections of the afterglows. We continued the monitoring campaign and obtained stringent upper limits on any re-brightening at the position of the optical afterglows up to 12 and 5 weeks after the bursts, respectively. The light-curves obtained based on this monitoring are shown in Fig. 1. For GRB 060505 we detected the optical afterglow at a single epoch. All subsequent observations resulted in deep upper limits. For GRB 060614 we followed the decay of the optical afterglow in the R-band up to four nights after the burst. In later observations no source was detected to deep limits (see also[14,15] for independent studies of this event). As seen in Fig. 1, the upper limits are far below the level seen in previous SNe, in particular previous SNe associated with long-duration GRBs[5-9]. For both GRBs our 3σ limit around the time of expected maximum of a SN component are 80-100 times fainter than SN 1998bw would have appeared. The very deep limits for GRB 060505 from May 23 and May 30 places a 3σ upper limits more than 250 times fainter than SN 1998bw at a similar time. Hence, any associated SN must have a peak magnitude in the R-band fainter than about −13.5.

A concern in any attempt to uncover a SN associated with a GRB is the presence of a poorly quantified level of extinction along the line of sight. In these cases however, we are fortunate that the levels of Galactic extinction in both directions are very low, *E(B–V)* = 0.02[16]. In the case of GRB 060505, our spatially resolved spectroscopy of the host galaxy allows us to use the Balmer emission line ratios to limit the dust obscuration at the location of the burst. The Balmer line ratio is consistent with no internal reddening. In the case of GRB 060614, the detection of the early afterglow in many bands, including the *Swift* UV bands UVW1 and UVW2[17], rules out significant obscuration of the source in the host galaxy and we conclude that there is no significant dust obscuration in either case (see also[15]).

Both GRBs were located in star-forming galaxies. The host galaxy of GRB 060505 has an absolute magnitude of about $M_B = -19.6$ and the spectrum displays the prominent emission lines typically seen in star-forming galaxies. The 2-dimensional spectrum shows that the host galaxy emission seen at the position of the afterglow is due to a compact H II region in a spiral arm of the host (see the supplementary material for details). We estimate a star-formation rate of 1 $M_\odot$ yr$^{-1}$ and a specific rate of about 4 $M_\odot$ yr$^{-1}$ (L/L*)$^{-1}$ (assuming $M^*_B = -21$). The host galaxy of GRB 060614 is significantly fainter with an absolute magnitude of about $M_B = -15.3$. This is one of the least luminous GRB host galaxies ever detected. We detect emission lines from hydrogen and oxygen and infer a star-formation rate of 0.014 $M_\odot$ yr$^{-1}$ (see supplementary material for details). The specific star-formation rate is 3 $M_\odot$ yr$^{-1}$ (L/L*)$^{-1}$. Sub-L* and star-forming host galaxies like these two are ubiquitous among long-duration GRB host galaxies[20]. For comparison, the specific star-formation rates of the 4 previously studied nearby ($z<0.2$) long-duration GRB host galaxies are 6, 7, 25 and 39 $M_\odot$ yr$^{-1}$ (L/L*)$^{-1}$ [21,22].

All spectroscopically confirmed SN-GRBs have peak magnitudes within half a magnitude of SN 1998bw[1,2], and all photometrically identified SN-GRBs have peak magnitudes within one and a half magnitudes of SN 1998bw[1,23,24]. For X-ray Flashes

(XRFs) there has been evidence that associated SNe span a somewhat wider range of luminosities[9,25-27], but still within the range of non GRB-selected Type Ic SNe. The available data on all Type Ic SNe (including SNe Ic unrelated to GRBs) show a distribution from roughly two magnitudes fainter in the *V* band than SN 1998bw to perhaps half a magnitude brighter[1,2,12]. The faintest SN Ic known is SN 1997ef (classified as a broad-line SN); but even a SN as faint as this would easily have been detected in our observations of GRB 060505 and GRB 060614 (SN 2002ap shown in Fig. 1 had a luminosity similar to that of SN 1997ef). Any SN associated with these two long-duration GRBs must therefore have been not only fainter than any SN previously associated with a GRB or XRF but also substantially fainter than any non-GRB related SN Ic seen to date.

The non-appearance of a SN in these cases is a surprise and indicates that we have uncovered GRBs with quite different properties from those studied previously. It is possible that the origin of these bursts lies in one of the many SN-less GRB progenitors suggested prior to the definitive association between GRBs and SNe. Also, in a variant of the the original collapsar model "fallback"-formed black holes or progenitors with relatively low angular momentum could produce SN-less GRBs[28,29,30]. Of the six long-duration GRBs or XRFs known to be at low redshift ($z < 0.2$), two now have no associated SN, so the fraction of SN-less GRBs could be substantial.

**Acknowledgements** We acknowledge benefits from collaboration within the EU FP5 Research Training Network "Gamma-Ray Bursts: An Enigma and a Tool". The Dark Cosmology Centre is funded by the DNRF. The observations presented in this article has been obtained from the ESO La Silla-Paranal observatory and from the Gemini Observatory. The Gemini Observatory is operated by the Association of Universities for Research in Astronomy, Inc., under a cooperative agreement with the NSF on behalf of the Gemini partnership: the National Science Foundation (United States), the Particle Physics and Astronomy Research Council (United Kingdom), the National Research Council (Canada), CONICYT (Chile), the Australian Research Council (Australia), CNPq (Brazil), and CONICET (Argentina).

**Competing interests statement**  The authors declare that they have no competing financial interests.

**Correspondence** and requests for materials should be addressed to J.P.U.F. (jfynbo@dark-cosmology.dk).

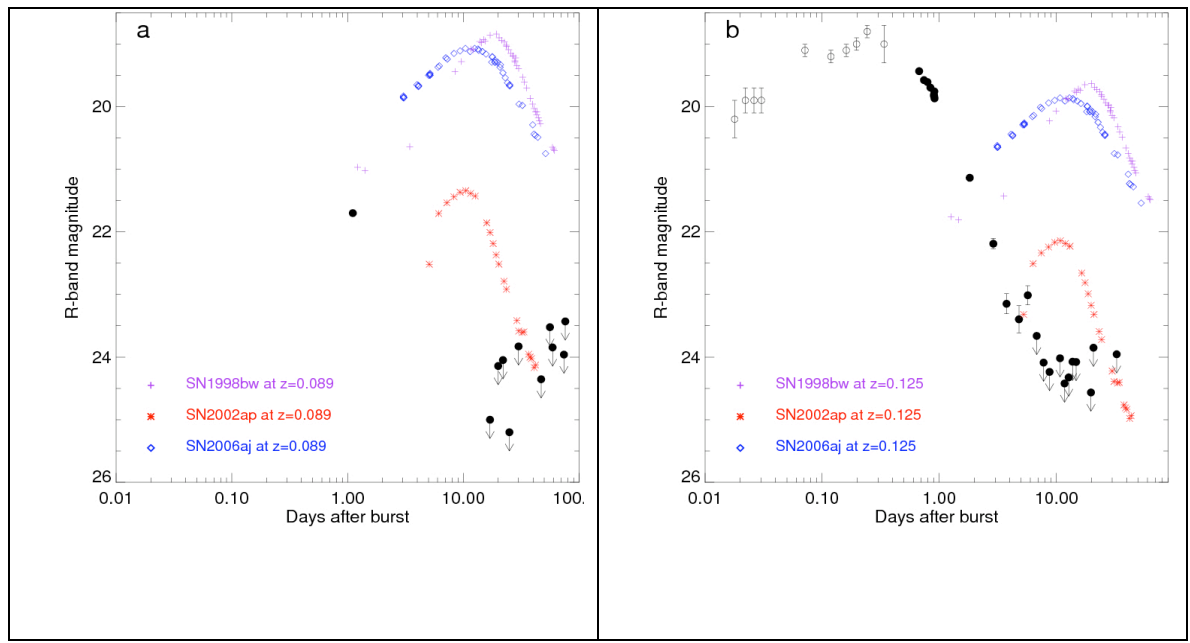

Fig 1. **No supernova associated with two nearby γ-ray bursts.** Lightcurves of SNe 1998bw, 2002ap and 2006aj, as they would have appeared at the redshift of GRB 060505 (**a**, left) and at the redshift of GRB060614 (**b,** right). We also plot our afterglow detection in both cases (filled circles) and subsequent 3σ upper limits. We conclude that neither GRB 060505 nor GRB060614 were associated with significant SN emission down to very faint limits hundreds of times less luminous than the archetypal SN 1998bw.

GRB 060505 was monitored with the D1.5m, the VLT and the Keck telescopes over a period of 12 weeks, which allowed us to detect the afterglow on the first night after the burst and thereafter obtain very strict upper limits on associated SN emission. GRB 060614 was observed with the D1.5m almost every night up to 5 weeks after the burst, which enabled us to monitor the decay of the optical afterglow, and finally to detect the host galaxy. For GRB 060614 the early light curve is populated with data reported in GCNs[17] (open circles).

These SN-less GRBs share no obvious characteristics in their prompt emission. GRB 060505 was one of the least luminous bursts discovered with *Swift* with an isotropic-equivalent energy release of $1.2 \times 10^{49}$ erg. It had a relatively short duration, and was single-peaked with a very faint afterglow. GRB 060614 was about a hundred times more luminous with an isotropic-equivalent energy release of $8.9 \times 10^{50}$ erg, and it showed strong spectral evolution. Its optical afterglow brightened for the first day[18], reminiscent of GRB 970508[19].

In the supplementary material we provide further details on the observations and data analysis.

# Supplementary material for "No supernovae associated with two long-duration gamma ray bursts"

## 1. Supplementary tables

In Table 1 and Table 2 we provide the observing logs. Most of the data were collected at the Danish 1.5m Telescope at La Silla in Chile (D1.5m) equipped with the Danish Faint Object Spectrograph and Camera (DFOSC). For GRB 060505 additional data were collected at the European Southern Observatory (ESO) Very Large Telescope using the FOcal Reducer/low dispersion Spectrographs (FORS1 and FORS2) and at the Keck telescope equipped with the Low Resolution and Imaging Spectrograph (LRIS). For GRB060614 additional data were obtained with the GEMINI Multi-Object Spectrograph (GMOS) instrument on GEMINI. The data were reduced using standard procedures for bias subtraction and flat fielding.

| Epoch | Exposure time (sec) | Telescope/Instrument | R-band magnitude |
|---|---|---|---|
| May 6.4 | 300 | VLT/FORS1 | 21.7 |
| May 23.3 | 120 | VLT/FORS2 | >25.0 |
| May 23.3 | 5530 | VLT/FORS2 | Spectrum |
| May 26.4 | 6000 | D1.5m/DFOSC | >24.1 |
| May 28.4 | 3000 | D1.5m/DFOSC | >24.0 |

| Epoch | Exposure time (sec) | Telescope/Instrument | R-band magnitude |
|---|---|---|---|
| May 30.6 | 300 | Keck/LRIS | >25.2 |
| June 5.4 | 3600 | D1.5m/DFOSC | >23.8 |
| June 22.4 | 1800 | D1.5m/DFOSC | >24.4 |
| July 1.4 | 3600 | D1.5m/DFOSC | >23.5 |
| July 4.4 | 2700 | D1.5m/DFOSC | >23.8 |
| July 19.4 | 3600 | D1.5m/DFOSC | >23.9 |
| July 20.4 | 5400 | D1.5m/DFOSC | >23.4 |
| July 30.6 | 300 | Keck/LRIS | Template image |
| Sep 14.2 | 300 | VLT/FORS1 | Template image |

**Table 1**: Observing log for GRB 060505.

| Epoch | Exposure time (sec) | Telescope/Instrument | R-band magnitude |
|---|---|---|---|
| June 15.2056 | 300 | D1.5m/DFOSC | 19.43 |
| June 15.2727 | 300 | D1.5m/DFOSC | 19.56 |
| June 15.3251 | 300 | D1.5m/DFOSC | 19.61 |
| June 15.3708 | 300 | D1.5m/DFOSC | 19.70 |
| June 15.4321 | 300 | D1.5m/DFOSC | 19.81 |
| June 15.4359 | 300 | D1.5m/DFOSC | 19.83 |
| June 15.4396 | 300 | D1.5m/DFOSC | 19.76 |
| June 15.4436 | 300 | D1.5m/DFOSC | 19.87 |
| June 16.2839 | 2700 | D1.5m/DFOSC | 21.14 |
| June 17.4 | 2700 | D1.5m/DFOSC | 22.19 |
| June 18.2 | 2700 | D1.5m/DFOSC | 23.15 |
| June 19.2 | 6600 | D1.5m/DFOSC | 23.40 |
| June 20.2 | 2700 | D1.5m/DFOSC | 23.01 |
| June 21.2 | 5900 | D1.5m/DFOSC | >23.7 |
| June 22.2 | 2700 | D1.5m/DFOSC | >24.1 |
| June 23.2 | 8100 | D1.5m/DFOSC | >24.2 |
| June 25.2 | 3600 | D1.5m/DFOSC | >24.0 |
| June 26.2 | 3600 | D1.5m/DFOSC | >24.4 |
| June 27.2 | 7000 | D1.5m/DFOSC | >24.3 |
| June 28.4 | 2700 | D1.5m/DFOSC | >24.1 |
| June 29.4 | 3600 | D1.5m/DFOSC | >24.1 |
| July 4.4 | 2880 | D1.5m/DFOSC | >24.6 |

| July 5.4 | 3300 | D1.5m/DFOSC | >23.9 |
| July 17.4 | 6300 | D1.5m/DFOSC | >24.0 |
| July 29 | 3600 | GEMINI/GMOS | spectrum |
| Aug 23-24 | 11000 | D1.5m/DFOSC | $B_{host}$=23.58 |

**Table 2:** Observing log for GRB 060614.

## 2. Data analysis

To remove the contribution from the host galaxies, we have used the template subtraction technique as implemented in the software package ISIS[31]. An example is seen in Fig. 1.

## 3. Comparison SN lightcurves

The SN lightcurves for SN1998bw, SN2002ap, and SN2006aj are taken from[5,32,33].

## 4. The star-formation rate of the GRB 060614 host galaxy

GRB 060614 was observed with the GMOS imaging spectrometer on the Gemini 8-m on the night of 29 July 2006 (see Table 2). Spectra were then obtained of the host using the nod and shuffle technique. While the spectroscopy began under excellent conditions, the seeing rapidly deteriorated, and the observations were terminated by the operators due to the increasingly poor seeing after only 3600s of integration time had been obtained. No spectrophotometric standard was observed.

In spite of the observing difficulties, the Hα and [O III] 5007 lines are easily detected. We derive a redshift of 0.1255 ± 0.0001. The Hα emission line has an equivalent width of 18 ± 2 Å. Scaling this by the R-band magnitude of the host we find an Hα flux of ~1.75×10$^{39}$ ergs s$^{-1}$, which corresponds to a SFR of 0.014 $M_\odot$ yr$^{-1}$ [34].

## 5. Could GRB 060505 and GRB 060614 be short GRBs or at higher redshifts?

The duration of the prompt emission GRB 060505 and GRB 060614 is 4 s and 102 s, respectively. Hence, they are both well outside the classification of short bursts, i.e. duration less than 2 s (note also, that the typical long-duration GRBs 000301C at $z$=2.04 and 020602 at $z$=4.05 had durations of only 2 s and 7 s, respectively, so their rest-frame durations are significantly shorter than those of GRB 060505 and GRB 060614). Short GRBs have previously had accompanying SNe excluded[35,36]. Therefore, one may speculate if the two SN-less GRBs studied here could be extreme members of the class of progenitors responsible for short GRBs. Short GRBs have in some cases been found to be associated with older stellar populations than long-duration GRBs, and it is widely expected that they are predominantly caused by merging, compact objects[37]. In addition to their long duration, the facts that 1) GRB 060505 and GRB 060614 can be localized to star-forming galaxies, 2) in the case of GRB060505 even a star-forming region in a spiral arm of its host galaxy (see Fig. 2), and 3) in the case of GRB 060614 the afterglow is located around the half-light radius of its star-forming host[14] strongly suggests that the progenitors were massive stars.

Nevertheless, the evidence in this paper suggests we should be more open-minded about what causes short GRBs: absence of SNe has been one of the arguments for saying short GRBs are not massive stars. Now we have observed SN-less GRBs in star-forming regions, suggesting that a non-detection of a SN does not preclude a massive progenitor. The position of the GRB, i.e. in a star-forming region or in an older component, may be the only way to discriminate between merging compact objects and massive stars as progenitors. In fact, several host galaxies for short GRBs have been found to be as actively star-forming as some host galaxies of long-duration GRBs[38,39]. The GRB labels "long" and "short" have become synonymous with "massive stars" and "other progenitors". These distinctions may need to be relaxed.

Is has also been suggested that GRB 060505 and GRB 060614 could in fact be more distant bursts and that the proposed hosts are only foreground galaxies. In the case of GRB060505, it is very unlikely that the afterglow is superposed on a star-forming region in the spiral arm of a foreground galaxy if it was in fact more distant. In the case of GRB 060614 the impact parameter is about 0.5 arcsec, and the afterglow is located around the half-light radius of the proposed host galaxy[14]. The UVOT detection of the afterglow[17] places an upper limit on the redshift of about $z\approx1$. At such relatively low redshift the host galaxy should be detectable in the very deep HST images of the field, but no other galaxy than the proposed host galaxy is seen[14]. Therefore, it is very unlikely that GRB 060505 and GRB 060614 are at higher redshifts than $z=0.089$ and $z=0.125$, respectively.

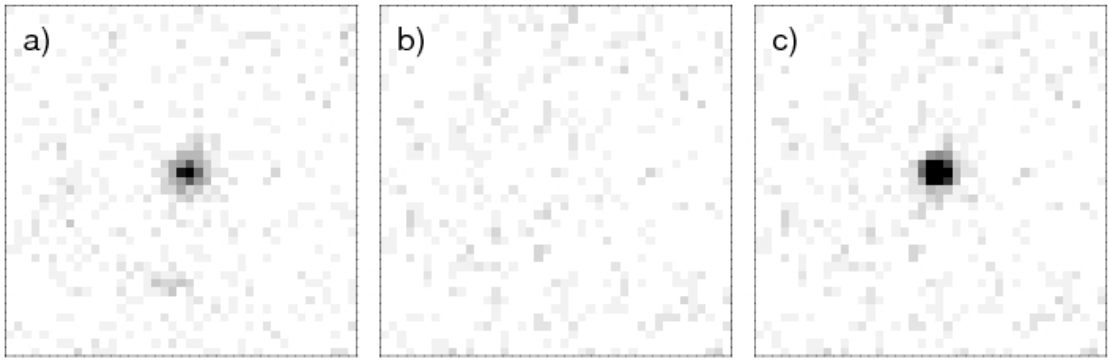

Fig 1. **Example of the image subtraction technique. a)** The field (16 times 16 arcsec$^2$) of GRB 060614 as observed from the Danish 1.5m telescope in the R band on June 25. The image shows the host galaxy of the burst, intrinsically a very faint galaxy, but clearly resolved in our image. **b**) The data after image subtraction analysis. There is no point source remaining in the data. **c**) This image shows how the very faint SN 2002ap would have appeared in our data.

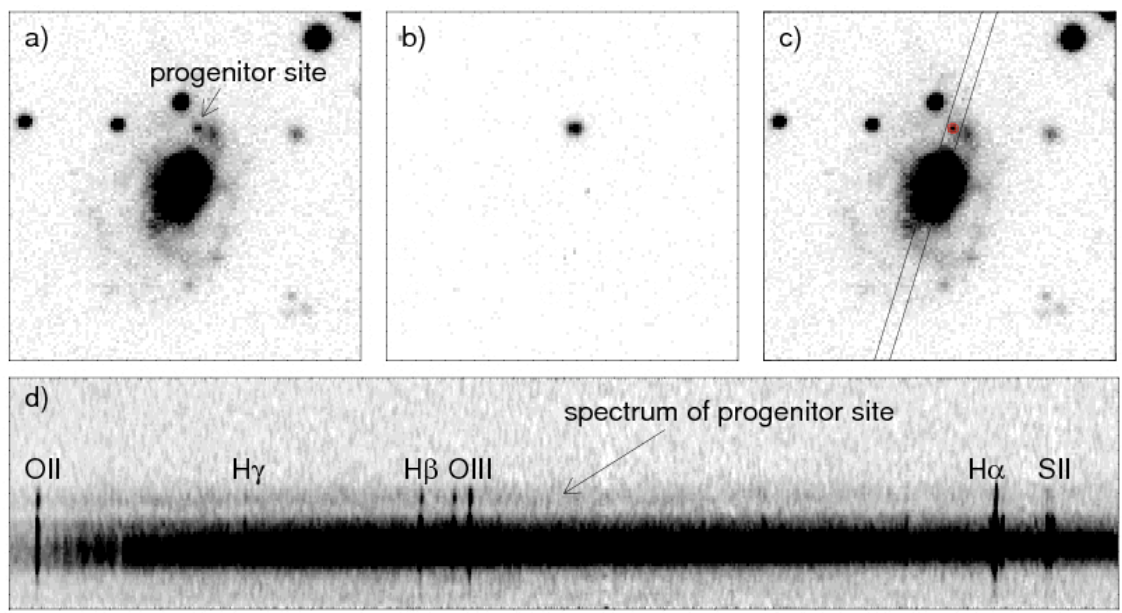

Fig 2. **The VLT image and spectrum of the GRB 060505 host galaxy. a)** The field (24 times 24 arcsec$^2$) of GRB 060505 as observed from the VLT in the R-band on Sep 14. The arrow marks the position where the optical afterglow was detected. The source seen at this position in the image is a compact star-forming region in which the progenitor of GRB 060505 was located. **b**) The result of subtracting the Sep 14 image seen in **a** from the May 6 image. Seen is the optical afterglow component alone. **c**) As **a**, but with the position of the optical afterglow marked with red contours and with the orientation of the slit for the May 23 spectrum indicated. The position of the afterglow is within the astrometric uncertainty of less than 0.05 arcsec coincident with the position of the compact star-forming region. **d**) The 2-dimensional optical spectrum obtained with VLT/FORS2 on May 23. As seen in **c** the slit covered the centre of the host galaxy and the location of GRB 060505. As seen in the spectrum, this site is indeed a bright star-forming region in the host galaxy and we hence have very strong evidence that the 060505 progenitor was a massive star. From the ratio of the Balmer line strengths we exclude dust extinction of more than a few tenths of a magnitude in the R-band.